\documentclass[aps,pre,showpacs,twocolumn]{revtex4-1}
\usepackage{amsmath} 
\usepackage{amssymb}
\usepackage{amsfonts}
\usepackage{makeidx}
\usepackage{graphicx}
\usepackage{epstopdf}
\include{epsf}
\usepackage{bm}

\newcommand{\bea}{\begin{eqnarray}}
\newcommand{\eea}{\end{eqnarray}}

\def\be{\begin{equation}}
\def\ee{\end{equation}}

\begin{document} 

\title{Solving the inverse Ising problem by mean-field methods in a clustered phase space with many states}

\author{Aur\'elien Decelle$^{1}$ and Federico Ricci-Tersenghi$^{2,3,4}$}
%\shortauthor{A. Decelle \etal}

%\pacs{02.50.Tt}{Inference methods}
%\pacs{02.30.Zz}{Inverse problems}
%\pacs{89.75.-k}{Complex systems}

%\institute{                    
%\inst{1} Dipartimento di Fisica, Universit\`a La Sapienza, Piazzale Aldo Moro 5, I-00185 Roma, Italy.\\
%\inst{2} INFN-Sezione di Roma 1, and CNR-IPCF, UOS di Roma. \\
%\inst{3} Laboratoire de Recherche en Informatique, TAO - INRIA, CNRS et Universit\'e Paris-Sud 11, B%\^at. 660, 91190 Gif-sur-Yvette, France
%}

\affiliation{
$^1$Laboratoire de Recherche en Informatique, TAO - INRIA, CNRS et Universit\'e Paris-Sud 11, B\^at. 660, 91190 Gif-sur-Yvette, France\\
$^2$Dipartimento di Fisica, Universit\`a La Sapienza, Piazzale Aldo Moro 5, I-00185 Roma, Italy. \\
$^3$Sezione di Roma 1, INFN, I-00185 Roma, Italy\\
$^4$Unit\`a di Roma, NANOTEC, CNR, I-00185 Roma, Italy}


\begin{abstract} 
In this work we explain how to properly use mean-field methods to solve the inverse Ising problem when the phase space is clustered, that is many states are present. The clustering of the phase space can occur for many reasons, e.g.\ when a system undergoes a phase transition, but also when data are collected in different regimes (e.g.\ quiescent and spiking regimes in neural networks). Mean-field methods for the inverse Ising problem are typically used without taking into account the eventual clustered structure of the input configurations and may lead to very poor inference (e.g.\ in the low temperature phase of the Curie-Weiss model). In this work we explain how to modify mean-field approaches when the phase space is clustered and we illustrate the effectiveness of the new method on different clustered structures (low temperature phases of Curie-Weiss and Hopfield models).
\end{abstract}

\maketitle

\section{Introduction}

The Ising inverse problem has been the subject of a large amount of works recently \cite{cocco2012adaptive,huang2013adaptive,mastromatteo2013beyond,ravikumar2010high,roudi2009ising}. Although this problem is known since many decades under the name of Boltzmann machine learning (BML), many recent applications and developments in different fields (e.g.\ biology \cite{morcos2011direct,bialek2012statistical,barton2013ising}, computer science \cite{hsieh2011sparse} and physics \cite{ricci2012bethe,nguyen2012mean,decelle2014pseudolikelihood}) have renewed the interest in studying such problems. The BML can be investigated under two very different approaches. In the first one, which concerns this work, a set of data is generated according to the Gibbs-Boltzmann measure of a generic Ising model. The input data for the inverse problem are therefore independent and distributed accordingly to the Boltzmann distribution of the system \cite{ackley1985learning}. In a second case, the data are generated according to a stochastic dynamical process which correlates configurations close in time, and this correlations in the input data are exploited in solving the inverse problem \cite{cocco2009neuronal}. In both cases, the traditional Bayesian approach consists in maximizing the likelihood function of the data. In this work, we focus on the first case which is commonly named ``static inverse Ising problem'' and is harder than the second case.

In the static case, maximizing the likelihood is a complicated task, because it directly depends on the partition function which is impossible to compute efficiently (in the general case, its complexity grows exponentially with the system size).  However, it is still possible to maximize the likelihood by the expectation-maximization method using a Monte Carlo (MC) algorithm and doing a Boltzmann learning procedure \cite{ackley1985learning}. The MC algorithm is used to evaluate the average value of the observables of the system (here the magnetizations and the correlations) and to update the value of the magnetic fields and the couplings by doing a gradient ascent. Yet, it is known that MC estimates do not converge quickly in many cases and may require many steps to obtain accurate mean values. It means that the MC algorithm should be run for a long time at each step of the BML procedure making the method quite slow.
For this reason, faster methods based on mean-field approximations are commonly used in practical applications.

In a recent work~\cite{nguyen2012mean} Nguyen and Berg have revisited the problem of finding a good mean-field (MF) approximation for the inverse Ising problem. It was already known that MF methods fail to provide a good couplings reconstruction at low temperatures even for ferromagnetic systems (see Fig.~\ref{fig:CW} for an example on a ferromagnet and \cite{mezard2009constraint} for an example on a MF spin glass). Worse than that, this problem in coupling reconstruction occurs also in cases where the MF approximation is exact in the thermodynamical limit (e.g.\ the Curie-Weiss model). This failure in reconstructing couplings in ferromagnetic systems can be understood by looking at the input configurations at low temperatures: below the ferromagnetic transition, indeed, configurations are clustered in two groups of respectively positive and negative magnetization. The naive MF (nMF) approximation is based on the self-consistency equations for the magnetizations, $m_i=\tanh(\beta\sum_j J_{ij} m_j)$, with $\beta$ being the inverse temperature, which have 3 solutions for $\beta>\beta_c$: it is well known that the $m_i=0$ solution is unphysical, while the two solutions with $m_i \neq 0$ are thermodynamically stable. However considering all the input configurations together the average magnetizations are zero by symmetry. Therefore, a naive use of MF equations infer the couplings using the unphysical $m_i=0$ fixed point, and lead to a very poor result.
Please notice that the same problem arises if one computes correlations in a naive way: using all input data connected correlations would not decay at long distance. Therefore, in order to use properly the nMF equations, it is mandatory to look at the two other solutions characterized by non-zero magnetization. These solutions arise naturally when considering the decomposition of the Gibbs-Boltzmann measure in the configuration space. 

The authors of Ref.~\cite{nguyen2012mean} consider the nMF equations for both states (of positive and negative magnetizations) at the same time. In this way they obtain an over-constrained system of linear equations to be solved. They manage to find a solution by using the pseudo-inverse of a matrix (see \cite{nguyen2012mean} for further details). We will see that this approach can be considerably simplified in the case of the Curie-Weiss (CW) model, and then generalized to models with many free-energy minima. In Ref.~\cite{nguyen2012mean} also the case of the Sherrington-Kirpatrick (SK) model is considered as a case study with a clustered phase space at low temperatures. We would like to emphasize, however, that the division in metastable states of the SK model is somehow problematic for this approach. The metastable states of the SK model in the glassy phase are highly non-trivial and therefore it is very difficult even to define them properly in a system of limited size. Therefore we claim that the inference algorithm of Ref.~\cite{nguyen2012mean} as well as the one presented in the present work are not suitable for this kind of models, for which more elaborate techniques (such as the pseudo-likelihood method \cite{besag1975statistical,wainwright2010AnnalHigh}) are required.

In the present work we show that couplings can be well inferred using nMF equations also in the low temperature phase if input configurations are previously clustered and the nMF inference algorithm is applied separately to data in each cluster. We show that our inference procedure based on solving the nMF or TAP equations inside each cluster \emph{separately} is much simpler than the method proposed in Ref.\cite{nguyen2012mean}, where self-consistency equations for each cluster need to be solved \emph{simultaneously}. Therefore the use of complicated numerical algorithms such as the pseudo-inverse is not necessary. In addition, we show that, at variance to what is claimed in Ref.~\cite{nguyen2012mean}, using the present inference procedure one does not estimate wrongly the magnetic fields. It is worth mentioning that, when using one of the MF fixed points with $m_i \neq 0$, a spurious magnetic field unavoidably appears due to errors on the inferred couplings. However  this magnetic field is very small and decreases when increasing the number of input data.

In order to prove that our method is very efficient we apply it to different kind of models. First we show that in the CW model the results are as good as those from more elaborated methods, like the pseudo-likelihood method. Then we focus on the Hopfield model where the number of different free-energy minima can be controlled and made larger. We show that it is possible to improve the results on the inference process by clustering the set of input configurations and to infer the right number of clusters to be used. We should mention that a previous attempt to infer the couplings in the (sparse) Hopfield model from data collected in a single state was done in \cite{braunstein2011inference}. However, in that work, the interaction network was assumed to be known and only couplings intensities were inferred, so a direct comparison with our results is not possible.

\section{Problem definition and inference algorithms}
\label{sec:model}

In the static inverse Ising problem one aims at inferring the value of the couplings between the variables and the eventual magnetic fields, given a set of $M$ equilibrium configurations.
More precisely we consider an Ising model with $N$ spins defined by the Hamiltonian
\begin{equation}
\mathcal{H}(\underline{s})=-\sum_{i<j} J_{ij} s_i s_j - \sum_i h_i s_i\;,
\end{equation}
where $i,j=1,...,N$. In the static case, the inference process is done by using input data distributed according to the Gibbs-Boltzmann measure
\begin{equation}
P_{\rm GB}(\underline{s})=\frac{e^{-\beta \mathcal{H}(\underline{s})}}{Z} \text{ where } Z=\sum_{ \underline{s} } e^{-\beta \mathcal{H}(\underline{s})}
\end{equation}
We remind here that the $M$ sampled configurations are assumed to be independent.

In the following we will consider two different families of inference methods. For mean-field methods, we shall consider the average magnetizations and correlations of the data
\begin{eqnarray}
  \bar{m}_i &=& \frac{1}{M} \sum_{a=1}^M s_i^{a} \\
  \bar{c}_{ij} &=& \frac{1}{M} \sum_{a=1}^M s_i^{a} s_j^{a}
\end{eqnarray}
These observables are the only information needed to infer the parameters of the models when using mean-field methods.
We will also consider the pseudo-likelihood methods for which the entire sampled configurations $\{s_i^a\}$ are needed. Let us now describe how these methods work and how we will used them in the context of a clustered phase space.

We first consider the naive mean-field approach where the equations can be simply derived by considering the solution of the Curie-Weiss model (where $J_{ij}=1/N$). For this model, the magnetisations and the correlations are given by
\begin{eqnarray}
	m_i &=& \tanh\left(\beta\sum_{j\neq i} J_{ij} m_j + \beta h_i\right) \label{eq:nMF_m} \\ 
	c_{ik} \equiv \frac{\partial m_i}{\partial h_k} &=& \beta  (1-m_i^2)\left( \sum_{j \neq i} J_{ij} c_{jk} +\delta_{ik} \right) \label{eq:nMF_c}
\end{eqnarray}
By inverting eq.~(\ref{eq:nMF_c}) we can reconstruct directly the couplings $J^*_{ij}$. Then, by using the $J^*$ and eq.~(\ref{eq:nMF_m}) we can infer the magnetic fields $h^*_i$
\begin{eqnarray}
	J^*_{ij} &=& -(c^{-1})_{ij} + \frac{\delta_{ij}}{\beta(1-m_i^2)} \label{eq:nMF_J} \\
	h^*_i &=& \beta^{-1}\bigg[{\rm atanh}(m^i) - \sum_{j\neq i} J^*_{ij} m_j\bigg] \label{eq:nMF_h}
\end{eqnarray}
We refer to this method as nMF in the rest of the article.

A second approximation commonly used is to consider the pseudo-likelihood method (PLM).  PLM is based on the maximization of the marginals probability of one spin $s_i$ given that the rest of the spins are fixed: $p(s_i|s_{j \backslash i})$. The PLM consists in maximizing the sum of all the log-pseudo-likelihood~\cite{besag1975statistical,wainwright2010AnnalHigh}
\begin{equation}
  \mathcal{P}\mathcal{L}= \frac{1}{NM} \sum_{i,a} \log(p(s_i^a|s^a_{j \backslash i}))
\end{equation}
In this method, we need to have access to all the configurations $\{s_i^a\}$. The advantage of this method is that it deals also with high order correlations and thus provides much better performances on finite dimensional systems \cite{ekeberg2013improved,decelle2014pseudolikelihood}, but it also can handle directly clustered phase space. Moreover it has a polynomial complexity at variance to using the true likelihood of the data.

\subsection{Clustering methods and inference with clustered phase space}
Here we describe the clustering algorithms that we use to divide configurations in clusters before applying the nMF method. These clustering algorithms group configurations together based on their distances: configurations are put in the same group if they are ``close'' enough and ``far'' from the other clusters, where the concepts of ``close'' and ``far'' usually need to be determined in a self-consistent way. We use the Hamming distance defined by $d_{ab}=1/(4N)\sum_i(s_i^a - s_i^b)^2$. In the present work we use two different clustering methods. First we consider the soft $K$-means clustering \cite{mackay2003information}. This method clusterizes the space of configurations by assigning each configuration to the closest of the $k$ centers ``softly'' (a configuration is assign to a center with a given probability). Then the position of the $k$ centers is updated accordingly to the position of the configurations inside each cluster. The procedure is repeated until convergence. This method is very fast, the complexity scale as $\mathcal{O}(M)$, but the results can depend strongly on the initial conditions (i.e.\ on how the $k$ centers are chosen at the beginning).

A second method is based on density clustering. The density clustering algorithm we consider~\cite{rodriguez2014clustering} first defines the density around each point. In our case the density is the number of configurations within a given range. Then, each data point is associated to its closest neighbor with higher density. This process naturally separates the phase space into a number of clusters which depend on the range used for defining the neighborhoods. Therefore by using this algorithm we do not need to specify the number of clusters. Thus this second clustering algorithm has the advantage of finding by itself the number of clusters. It suffers however of a larger complexity, scaling as $\mathcal{O}(M^2)$.

After clustering the configurations we have to use them properly to infer the parameters of the model. We define the observables of the $k^{\rm th}$ cluster by 
\begin{eqnarray}
\bar{m}_i^{(k)} &=& \frac{1}{M_k} \sum_{a \in \mathcal{C}_k} s_i^a \\
\bar{c}_{ij}^{(k)} &=& \frac{1}{M_k} \sum_{a \in \mathcal{C}_k} s_i^a s_j^a
\end{eqnarray}
where $\mathcal{C}_k$ is the set of indices of configurations belonging to the $k$-th cluster and $M_k=|\mathcal{C}_k|$. We now apply the nMF equations separately for each cluster and obtain a different estimate of the parameters for each cluster $J_{ij}^{(k)}$. Finally, to obtain the best estimate for the couplings we take the weighted average of all the different estimates
\begin{equation}
  J_{ij}^{*} = \frac{1}{M} \sum_k M_k J_{ij}^{(k)}
\end{equation}
To estimate the magnetic field, we first compute them within each cluster: $h_i^{(k)}$ is obtained from eq.~(\ref{eq:nMF_h}) with the estimates $J_{ij}^{(k)}$. The final estimate for the magnetic fields is again given by the weighted average over the clusters
\begin{equation}
h_i^{*}=\frac{1}{M} \sum_k M_k h_i^{(k)}
\label{eq:h_final}
\end{equation}

\section{Results on the Curie-Weiss model}
\label{sec:methodCW}

The Curie-Weiss (CW) model is a fully connected ferromagnet with $J_{ij}=1/N$, $\forall i\neq j$. The model has a paramagnetic phase ($m_i=0$) at high temperature $\beta<\beta_c=1$ and a ferromagnetic phase ($m_i\neq 0$) above $\beta_c$. In the ferromagnetic phase, two states of positive and negative magnetizations coexist. In the limit of very large system sizes ($N \to\infty$) magnetizations and correlations can be computed analytically by eqs.~(\ref{eq:nMF_m}-\ref{eq:nMF_c}), which are exact up to $\mathcal{O}(1/N)$ corrections. It means that, by using eqs.~(\ref{eq:nMF_J}-\ref{eq:nMF_h}) one should obtain the best possible estimate of the parameters $J_{ij}$ and $h_i$, but in the ferromagnetic phase, only the solution with non-zero magnetization of the eq.~(\ref{eq:nMF_m}) should be considered (as discussed in the Introduction). We evaluate now how the following three inference algorithm perform in the estimate of couplings in the CW model: (i) the nMF method used naively, without clustering the configurations; (ii) the nMF method on configurations clustered using two clusters; (iii) the PLM on the original configurations.

\begin{figure}
  \centering
  \includegraphics[width=\columnwidth]{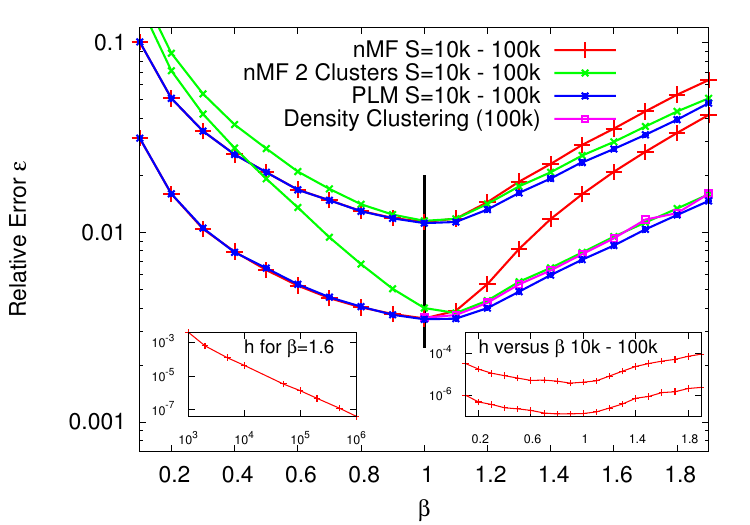}
  \caption{Inference of couplings in the CW model with $N=100$ and two different values of the number $M$ of input configurations. We see that the nMF method with all the input data is good only for $\beta<\beta_c=1$. For $\beta>\beta_c$ the phase space separates in 2 states and the nMF method with 2 clusters give much better performance (although it fails badly at high temperature). Inference methods, like PLM and nMF with density clustering, that take correctly into account the clustering of input configurations provide the best estimate in the entire temperature range, both above and below the transition temperature. In the left inset, we show how the inferred magnetic field at $\beta=1.6$ decreases when increasing the number of samples ($M\in[10^3,10^6]$) used for the inference process via nMF with $K$-means clustering and $K=2$. In the right inset the same inferred magnetic field is plotted versus $\beta$, for $M=10^4,10^5$.}
  \label{fig:CW}
\end{figure}

In Fig.~\ref{fig:CW} we report the error achieved by different methods in the temperature range $\beta \in [0.1,2]$ with $M=10^4$ and $M=10^5$ in inferring the couplings using the following definition

\begin{equation}
	\epsilon^2 = \frac{2}{N(N-1)} \sum_{i<j} \left( J_{ij}-J_{ij}^* \right)^2
\end{equation}

For $\beta<\beta_c$ the paramagnetic fixed point is correct and therefore the reconstruction achieve by nMF is the best possible. However, for $\beta>\beta_c$ the nMF error (red curves) suddenly raises, because the $m_i=0$ fixed point is no longer the physical one. On the contrary, using the nMF method on the data clustered with exactly 2 clusters (green curves), provides a small error in the ferromagnetic phase, but fails badly in the paramagnetic phase. The inference methods that provide the best estimate in the whole temperature range are the PLM (blue curves) and the nMF with data clustered via density clustering (purple curve), that automatically split the input data in one or two clusters, depending on symmetries in the input data.
It is worth stressing that these two methods have essentially the same error at any temperature: that is even the nMF approximation provides the best possible estimates if applied to properly clustered data.

In Fig.~\ref{fig:CW} we show results obtained with $M=10^4$ and $M=10^5$ in order to make evident whether the uncertainties in the couplings estimates are due to the noise in the input data or to an intrinsic limitation of the inference algorithm. For example deep in the ferromagnetic phase the nMF method has an error decreasing only slightly when $M$ increases, because the error is mainly due to a limitation of the method. On the contrary, PLM and nMF with properly clustered data provide a result whose uncertainty is mainly due to noise in the input data: indeed the error decreases as $1/\sqrt{M}$.

To confirm the correctness of the inference algorithm based on data clustering and nMF equations, we also looked a the inferred value of the magnetic field by using eqs.~(\ref{eq:nMF_h}) and (\ref{eq:h_final}). We see clearly in the insets of Fig.~\ref{fig:CW} that, in the low temperature phase, the clustering+nMF method does not predict any anomalously large magnetic field, thanks to the fact that, clustering the input data, we are actually using the magnetized solutions of eq.~(\ref{eq:nMF_h}). In our numerical experiments, we have found too large inferred magnetic fields only if either system size was too small or the input data were too noisy: in the former case the problem resides in the fact eq.~(\ref{eq:nMF_h}) is crudely approximate, while in the latter case it is a consequence of large errors in couplings reconstruction.

% . What we observe in this case is that the inferred magnetic fields in the low temperature phase are indeed larger when using the two states with respect to the case where the paramagnetic fixed point is used. This can easily be understood since, in the paramagnetic solution the magnetisation is equal to zero which tend to estimate a smaller residual magnetic field. However we have clearly observed that, by using the correct solution, the residual magnetic field decrease with the system size.

\section{Results on the Hopfield model}
\label{sec:results}

\begin{figure}[!h]
	\centering
	\includegraphics[width=\columnwidth]{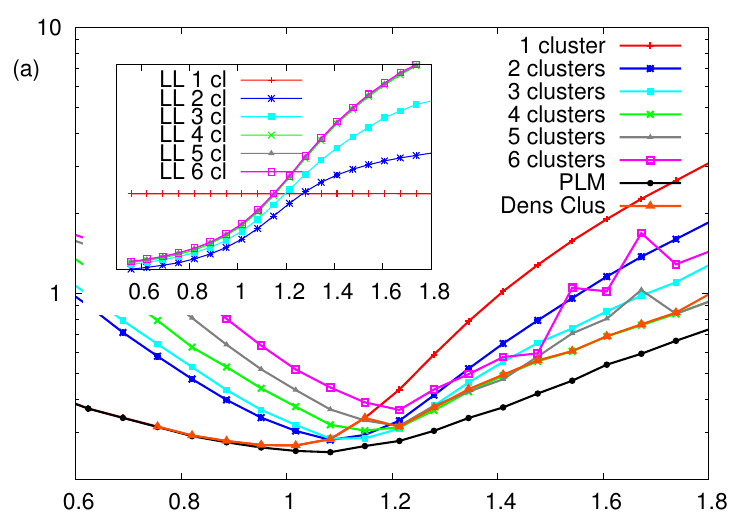}
	\includegraphics[width=\columnwidth]{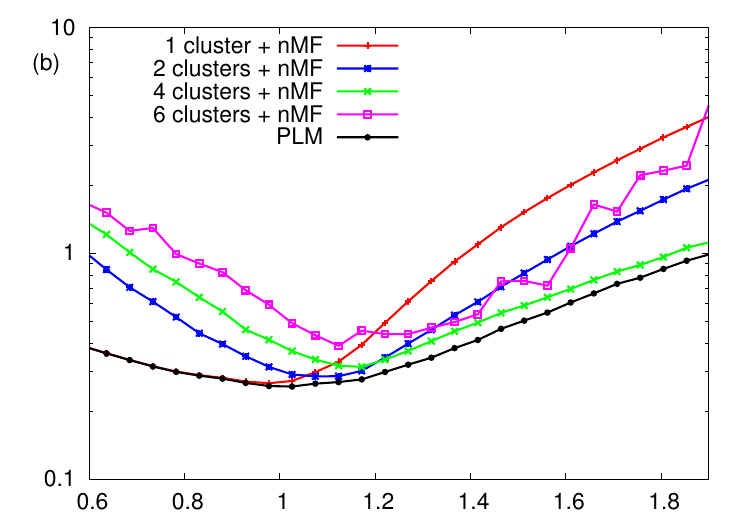}
	\includegraphics[width=\columnwidth]{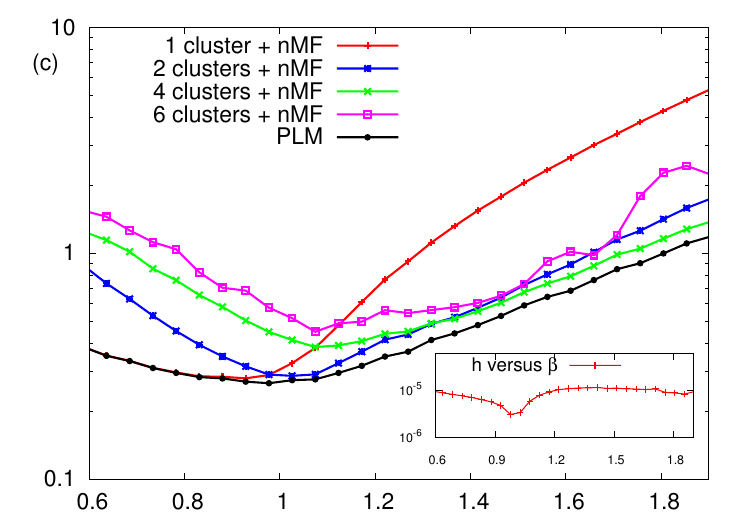}
	\caption{Main panels: errors in inferring couplings in Hopfield models with $P=2$ uncorrelated (top), correlated (center) and anti-correlated (bottom) patterns. The comparison is between MF methods with clustered data (either $K$-means or density clustering) and PLM. In the inset of the top panel, we show that the likelihood of the clustering algorithm suggests to take one cluster below $\beta_c\approx1.1$ and 4 clusters above $\beta_c$. In the inset of the bottom panel we show the magnetic field inferred by nMF+clustering, which is very small in both phases.}
	\label{fig:2clus}
\end{figure}

We now extend our analysis to a more complicated case by considering the Hopfield model. The Hopfield model has been introduced long time ago \cite{hopfield1982neural} to model neural networks: it is a fully-connected Ising model, whose couplings can be chosen such that the model free-energy has $2P$ different minima (that act has attractors for the pattern recovery dynamics). In some sense, the Hopfield model can be seen as a generalization of the Curie-Weiss model, which is indeed equivalent to the $P=1$ case.
We are interested in studying the inverse Ising problem in the Hopfield model, because configurations sampled at low temperature in the Hopfield model are typically clustered around the $2P$ free-energy minima:
consequently naive MF methods face even more severe limitations than in the low temperature phase of the CW model, and we want to study how much MF methods for the inverse Ising problem can be improved by clustering input configurations.

The Hamiltonian of the Hopfield model reads
\begin{equation}
  \mathcal{H}(\underline{s}) = -\frac{1}{N} \sum_{ij} \frac{1}{P} \sum_{\alpha=1}^P \xi_i^\alpha \xi_j^\alpha s_i s_j\;,
\end{equation}
where the $P$ patterns $\xi^\alpha$ identify the directions of the free-energy minima.
In the standard Hopfield model, the $\xi$s are drawn from the bimodal distribution, that is $\xi_i^\alpha= \pm 1$ with probability $1/2$ independently. In our study we also consider the case where the pattern $\xi$ are correlated by setting $10$\% of their components equal ($\xi_i^\alpha=\xi_i^\beta \quad \forall \alpha,\beta$), and anti-correlated (only when $P=2$) by setting $10$\% of their components in an opposite way ($\xi_i^1=-\xi_i^2$). This model presents a paramagnetic phase at high temperature, and an ordered phase at low temperature defined by the states around the patterns $\{\xi\}$ if the number of patterns is not too high \cite{amit1985storing}. The ordered phase is characterized by a Gibbs-Boltzmann measure clustered around one of the $2P$ available states (for a given $P$ there will be $2P$ stable states in the low temperature region due to the spin flip symmetry).

We show now our results on inferring the Hopfield couplings by using MF methods on clustered data. In Fig.~\ref{fig:2clus} we consider systems with $N=100$ spins, $P=2$ (therefore 4 states) in all the three possible cases (standard, correlated and anti-correlated patterns).
We observed that MF methods with the right number of clusters perform similarly to the PLM, which is at present the best possible algorithm to solve the inverse Ising problem. The right number of clusters can be obtained either by density clustering or by maximizing the likelihood of the clustering obtained by $K$-means (see panel (a) inset in Fig.~\ref{fig:2clus}).

As in the CW model, also for the Hopfield model the magnetic fields inferred by MF methods on clustered data are very small, and independent on the eventual long range order present in the model (see inset in panel (c) of Fig.~\ref{fig:2clus}).

\begin{figure}
	\centering
	\includegraphics[width=\columnwidth]{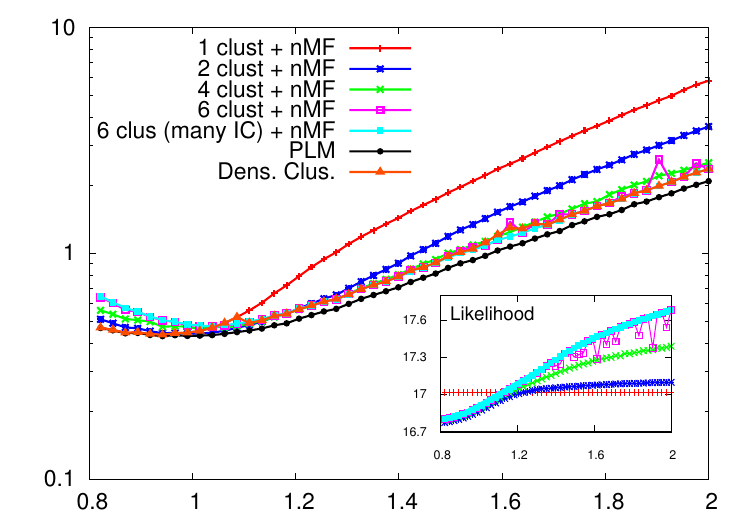}
	\caption{Errors on inferring couplings in the Hopfield model with 3 patterns (and thus 6 minima). We observe again that our algorithm, based on MF methods applied to clustered data, achieves its best performance when input data are split in 6 clusters. We also put for comparison the results obtained when the clustering is done many times with different initial conditions (label `many IC') and then we picked the clustering having the largest likelihood. In this case, the error matches the error obtained when putting each configuration in the correct cluster. We can see that our method performs its best at almost any $\beta$ value, but at few points where it is particularly difficult to find the best clustering. In the inset we see that likelihood maximization suggests to use 1 cluster for $\beta<\beta_c$ and 6 clusters for $\beta>\beta_c$.}
	\label{fig:6clus}
\end{figure}

In Fig.~\ref{fig:6clus} we show the results on inferring couplings of Hopfield models with $P=3$ patterns (and thus 6 free-energy minima). Again MF methods applied on input data clustered with the right number of clusters perform very similarly to PLM, and much better than standard MF methods applied directly to all input data.
It is worth noticing that the best result by the clustering+nMF algorithm as been obtained by running the clustering procedure several times with different initial conditions (data labeled 'many IC' Fig.~\ref{fig:6clus}) and then picking the clustering having the largest likelihood.
This is expected since a clustering algorithm as $K$-means is not very stable for large $K$ and its outcome strongly depends on the initial condition.

Let us finally discuss the time complexity of the three algorithms we have used: PLM, $K$-means+nMF and dens.clus.+nMF.
Regarding the system size dependence, all three algorithms have a time complexity $\mathcal{O}(N^3)$, either because of the inversion of a $N\times N$ matrix in nMF methods, either because of the computation of the gradient of the pseudo-likelihood (PL), which is $\mathcal{O}(N)$, in a space of $\mathcal{O}(N^2)$ variables.
Their dependence on the number $M$ of input configurations is different: PLM is linear in $M$, but the search for the maximum of the PL, requires to compute PL and its derivatives many times; $K$-means is linear in $M$, but often a search for the optimal clustering requires to run it with many different initial conditions; density clustering is $\mathcal{O}(M^2)$, so, although it provides a robust result, it is impracticable when the number of samples is very high (however we are aware that the authors of Ref.~\cite{rodriguez2014clustering} are developing a faster version of the density clustering algorithm). Therefore, nMF methods are always faster with a total complexity of $\mathcal{O}(KMN + N^3)$ whereas PLM is $\mathcal{O}(M N^3)$.
In practice, we observe it is better to use PLM when the number $M$ of input configurations is small since it gives in general better estimates of the reconstructed couplings. When $M$ becomes large, nMF with $K$-means clustering is clearly recommended since PLM would be affected by the large number of samples.

\section{Conclusions}
\label{sec:ccl}

In this work we have presented a very simple way to make mean field approximations to the inverse Ising problem effective also in the low temperature phase, where symmetries get usually broken and, correspondingly, input data get clustered.
The idea is to cluster the input data and to apply mean-field methods to each data cluster.
We have tested this clustering+nMF algorithm on the Curie-Weiss and Hopfield models, comparing results with the most sophisticated and state-of-the-art pseudo-likelihood method.

Results are very promising and redeem mean-field approximations to inverse problems, even in those cases where the structure of the input data is such that a straightforward application of mean-field methods would be ineffective.

The natural follow-up to this work is application of clustering+nMF methods to inverse problems based on real data. It is worth remembering that often in solving inverse problems based on real and noisy data, the robustness of simple MF methods is more valuable than the putative higher accuracy of more sophisticated methods: see e.g. the case of inferring protein contacts \cite{morcos2011direct}. 
%AD added for revision
It is also worth mentioning cases where the data can be naturally divided in two or more classes, exhibiting different statistical properties, but this is usually not taken into account when estimating model parameters. For example Ref.~\cite{panas2015sloppiness} presents an impressively detailed analysis of neuronal spiking patterns. Nonetheless the data belonging to two different regimes (quiescent and spiking) are merged together before doing the analysis according to mean-field approximation. The application of the method presented in this work is likely to improve inference and reduce errors.
Finally, the numerous recent studies on pattern recognition using neural network might also benefit from an approach dealing with clusters. In those systems it is quite common to deal with many basins of attraction that are used to improve the neural network efficiency. Mean-field techniques would be more than welcome since methods such as PLM cannot deal with the large dataset size (particularly since the average over all samples has to be done at each step of the algorithm).

From this point of view, enlarging the range of applicability of MF methods by data clustering is certainly very useful and maybe better than developing higher order approximations (that strongly depends on the model used to describe the data).

\bibliography{Bib}

\end{document}